# ONTOLOGY-BASED RECOMMENDER SYSTEM OF ECONOMIC ARTICLES

David Werner, Christophe Cruz and Christophe Nicolle
*LE2I Laboratory, UMR CNRS 5158*
*BP 47870, 21078 Dijon Cedex, France*
*{david.werner, christophe.cruz, cnicolle}@u-bourgogne.fr*

Abstract: Decision makers need economical information to drive their decisions. The Company Actualis SARL is specialized in the production and distribution of a press review about French regional economic actors. This economic review represents for a client a prospecting tool on partners and competitors. To reduce the overload of useless information, the company is moving towards a customized review for each customer. Three issues appear to achieve this goal. First, how to identify the elements in the text in order to extract objects that match with the recommendation's criteria presented? Second, How to define the structure of these objects, relationships and articles in order to provide a source of knowledge usable by the extraction process to produce new knowledge from articles? The latter issue is the feedback on customer experience to identify the quality of distributed information in real-time and to improve the relevance of the recommendations. This paper presents a new type of recommendation based on the semantic description of both articles and user profile.

## 1 INTRODUCTION

The decision-making process in the economic field requires the centralization and the consummation of a large amount of information. This aims at keeping abreast with current market trends. The Company Actualis SARL is specialized in the production and distribution of press reviews about French regional economic actors. This economic review represents for a client a prospecting tool on partners and competitors. The reviews sent are the same for each customer, which does not necessarily correspond to its needs. From the result an opinion surveys on clients and the knowledge on company's business from the company Actualis SARL, criteria for relevant review customization were identified. These criteria are economic events, economic sectors, major transverse projects, temporal and localization data about each element underlined. To reduce the overload of useless information, the company is moving towards a customized review for each customer. To achieve these goals, a recommender system is being developed (e.g. fig 1.). This system is regularly supplied with articles by the company librarians. It produces a magazine per customer composed of a subset of daily produced articles according to the client's profile. This system is composed by a couple of layers. The first, the Intelligence Layer aim to manage information extraction tasks, it contains several mechanisms. The second, Semantic Layer is composed of ontologies. This Layer allows to manage general or field specialized knowledge, to model profiles and articles representation. However, three issues appear to achieve this goal. The first challenge is to identify the elements in the text in order to extract objects that match with the recommendation's criteria presented previously. Additionally, the links between these objects have to be also extracted from the articles, because they represent valuable information. The second issue lies in the definition of the structure of these objects, relationships and articles in order to provide a source of knowledge usable by the extraction process to produce new knowledge from articles. This recommended system permits the economic watch on potential clients and eventually to send appropriate alerts to customers about important and new information or knowldege. The latter issue is the feedback on customer experience to identify the quality of distributed information in real-time and to improve the relevance of the recommendations. This is materialized by the evolution of the customer profile in real time through the review reading, article by article.

The ontology plays a center role because it is used to model the knowledge of the domain, and to drive the extraction process and the recommendation process.

The following section presents an overview on knowledge extraction, which is related to the first issue. Section 3 presents the recommender systems which allows us to identify the relevant architecture of the recommended system for the articles and knowledge recommendation. Section 4 focuses on the architecture of the recommendation system which is set up especially on the four ontologies. These ontologies will be used by all business processes, which are the knowledge extraction process, the article indexing process, the article annotation process, the recommendation process.

## 2 KNOWLEDGE EXTRACTION

In this section, we present an overview of information extraction systems based on ontologies, and we focus on their main functions and architecture. The information Extraction (IE) is intended to extract specific data elements as entities, relationships or events from a set of textual records. The approach generally used is based on the use of rules, patterns applied to texts by transducers or finite state automata. The presentation of information retrieval architecture is available in (Daya C. Wimalasuriya, 2009). KIM (B. Popov and all, 2003) can be seen as an OBIE (Ontology-Based Information Extraction) system, it used the KIMO (www.ontotext.com/) ontology the predecessor of PROTON (proton.semanticweb.org) to manage the necessary knowledge for the annotation task. During text analyses, patterns and gazetteers based approach is used to extract information, like organizations names, persons or dates. New information extracted, detected by patterns are used to populate the ontology. Information extracted during previous text analyses are used to perform future text analyses. Texts and ontologies can be both seen as sources of knowledge, and they are complementary, because each one providing information to the other.

On the one hand, the text is a source of information to ontologies. The first case consists in reaching the automation of the ontology creation regarding a domain defined by a set of documents. Therefore, treatment aims to detect in the text, key concepts and their properties and relationships. This task is called ontology learning. The second case focuses on the automation of the individual detection, properties and relationships in the documents in order to supply the knowledge base. This task is called ontology population.

On the other hand, where ontologies are used to provide information to the texts, the idea is to describe the knowledge contained in texts with known information contained the knowledge bases. Groups of words in texts can be labeled according to the ontology to highlight instances of concepts or relationships. Therefore, ontologies seem to be the fundamental tool to meet our needs of indexing and annotating of the economic articles. In addition, OBIE is a tool able to index, which is a second important tool for our purpose. Moreover, by its nature, the ontology contains the relevant criteria of recommendation. Consequently, after the indexing and annotating tasks executed over articles, the ontology provides a good support for our recommender system. The following section presents a brief state of the art of recommender systems focuses on the needs of our recommender system.

## 3 RECOMMENDER SYSTEMS

Recommender systems (RS) are tools and pieces of software, which aims are to provide suggested items to users (Burk, R., 2007). The simplest task of RS is the ranking of items (e.g. Books, CDs, Travels, and so on); it tries to predict the importance each item for the user to ordering each other. The computing task can be based on explicit user's preferences, (e.g. the user give a score to product), or implicit, preferences are inferred by the system, depending on user comportment (e.g. choices of browsing on an online shopping site).

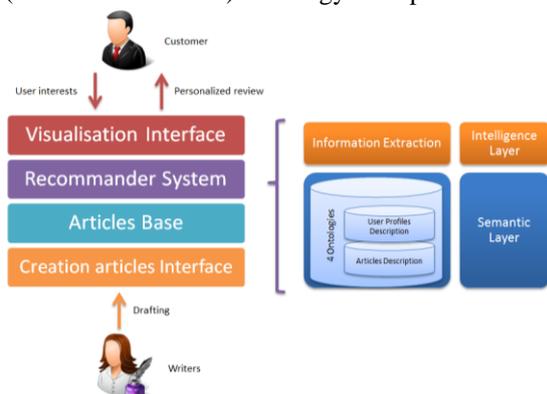

Figure 1. The architecture of our recommender system

The increasing need of right information in a right time requires the development of recommender systems to drive users. An overview and taxonomy of the different kinds of systems can be found in (Burk, R., 2007), both main approaches are distinguished, content-based and collaborating filtering based.

- Content-based approach: This kind of RS tries to recommend similar items of already known relevant items, or compare items and users profiles. All items and profiles are characterized by different kind of attributes. I.e. if a user evaluates positively a thriller book, then the system trends to recommend other thriller books.
- Collaborating filtering based approach: This Kind of RS tries to predict automatically the interest of a user with the help of taste information from many other users. An implantation of this approach (Schafer, Frankowski, Herlocker, Sen 2007) consist in recommend items appreciated by users with close profiles. This similarity of profiles is computed in relation to previous item's grade by each profile. I.e. if close users (users with profiles close to user profile) mostly enjoy « le discours de la methode », then the system trends to recommend this book to users.

The approach developed here for the recommendation is based on the information, and the knowledge extracted from each economical article. Each article contains a set of information pieces, and each of these pieces is used like an attribute characterizing the articles. The set of properties and the set of known possible values of each allow the definition of a structured representation of each article. In most content-based filtering systems, items are described by textual properties, by words. Some properties of natural language create trouble in the use of the word in matching task, like polysemy (matching profile with no really relevant items) or synonymy (no matching profile with yet relevant items). In order to remedy, semantic-based techniques were developed.

- Keyword vectors: Most of RS use simple methods like key words or vector space models (VSM), with or without moderation process like TF-IDF (Salton, 1989). VSM is spatial representations of documents. Each text is represented by an n-dimensioned vector; n is the number of words selected by generally statistical processes to representing the document. The similarity between tow vectors (item's vector/item's vector or item's vector/profile's vector) can be measured with the computation of cosine similarity.
- Semantics: There are lots of strategies to introduce semantics in the task of recommendation. These strategies are generally based on ontologies. Different kinds of ontologies are used, like Wordnet (A lots of different words can be used to referring the same concept in natural language) or like domain specifics ontologies (to create a structured and controlled representation of domain, which can be used to describe items).

Vector realizes the qualification of articles and profiles. A set of criteria corresponding to information extracted from articles forms the vector (i.e. locations, economic events, transversal projects and economic sectors).

Moreover, the extraction of relevant knowledge extends the use of the knowledge base where the knowledge drives the recommendation. Thus, not only articles are recommended but also the knowledge contains in a set of articles. From this knowledge, articles can be combined to present a global view of a specific domain.

## 4 THE ECONOMIC RECOMMENDER SYSTEM

The section deals with the component of the architecture and the aims of these components.

### 4.1 The architecture

Figure 2 depicts the architecture of processes of our recommender system. Annotation and indexing processes bind articles and knowledge using the OBIE system. These processes make possible the creation and the management of the semantic descriptions of articles. Next, profiling process makes the semantic representation of user's profiles with vectors of concepts. These vectors link users and knowledge base.

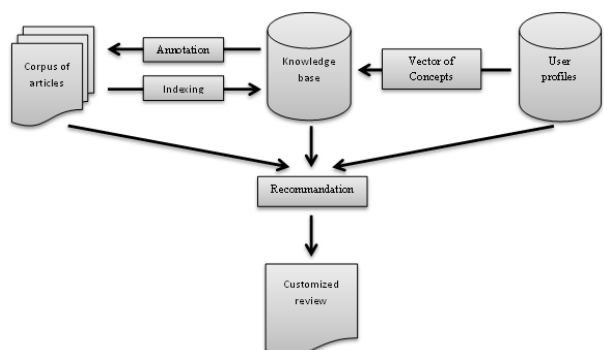

Figure 2. The architecture overview

The user profile can be determined in various ways, including active and passive feedback, allowing us to know what information is relevant to each user.
The last process, generate a customized review for each user made of articles according to his profile.
Profiles and knowledge extracted from articles are described using an ontology. Using this ontology articles can be recommended.

### 4.2 The knowledge representation

Description logics (DLs) are a family of logics that are decidable fragments of first-order logic with attractive and well-understood computational properties. DLs have been in use for over two decades to formalize knowledge and notably quality ontologies. Ontology languages like OWL DL and OWL Lite semantics are based on DLs (Horrocks, 2009). For example, OWL DL corresponds to the SHOIN (D) description logic, while OWL 2 corresponds to the SROIQ(D) logic (Hitzler and al, 2009). Our work deals with OWL DL ontologies so we chose the SHOIN(D) expressivity level to formalize ontology inconsistency. In DL, a distinction is drawn between the so-called TBox (terminological box) and the ABox (assertional box) (Gruber, 1993). In general, the TBox contains sentences describing concept hierarchies (i.e., relations between concepts) while the ABox contains ground sentences stating where in the hierarchy individuals belong (i.e., relations between individuals and concepts). In OWL DL ontologies, TBox corresponds to the intension and ABox to the extension. Ontologies are knowledge representation, a description understandable bye the machine. The indexing task based on an ontology allow the definition of the knowledge structure which limits the ambiguities inherent in the use of simple words. Ontology is a representation of a context, which permits a formal interpretation of the information contained herein. Our knowledge base consists of four ontologies: The upper-level ontology, the domain ontology, the lexical resource ontology and the corpus ontology. The first two ontologies intend to distinguish the knowledge specific to an application domain (domain ontology) from those which transcend all areas (upper-level ontology). The Lexical resource ontology is inspired by PROTONS. It is used in the management of objects required by NLP tools. These tools are used to perform the information extraction task. The corpus ontology manages the items to be indexed. In our case these are articles.
This model aims to make the system less dependent on a given area. It allows us to change the domain ontology in order to move from one area to another. The ability to switch the domain ontology with another one makes our system flexible.

## 5 CONCLUSIONS

This work presents a new approach for recommender systems based on a set of four ontologies. This generic proposal has been applied in the field of economic reviews. The system built aims at providing to company's customers a set of economic articles, which contain information relevant to their business needs.
In the work presented, the bias is to propose the recommendations based on the knowledge included in the articles. Information extraction systems were presented including those based on ontologies. They allow both to evolve the index (populating the knowledge base) and to index articles.

## ACKNOWLEDGEMENTS

This project if founded by the company Actualis SARL and the financing CIFRE research grant from the French agency ANRT.